*Full Length Research Paper*

# Economic valuation of tourism of the Sundarban Mangroves, Bangladesh

Mohammad Nur Nobi[1], A. H. M. Raihan Sarker[2*], Biswajit Nath[3], Eivin Røskaft[4], Ma Suza[5] and Paul Kvinta[6]

[1]Department of Economics, University of Chittagong, Chittagong - 4331, Bangladesh.
[2]Institute of Forestry and Environmental Sciences, University of Chittagong, Chittagong- 4331, Bangladesh.
[3]Department of Geography and Environmental Studies, University of Chittagong, Chittagong- 4331, Bangladesh.
[4]Department of Biology, Norwegian University of Science and Technology, NTNU, Trondheim #7491, Norway.
[5]Department of Social Sciences, Wageningen University, PO Box 8130, 6700E Wageningen, The Netherlands.
[6]573 Cameron Street SE, Atlanta, Georgia 30312, United States.

The Sundarban Reserve Forest (SRF) of Bangladesh provides tourism services to local and international visitors. Indeed, tourism is one of the major ecosystem services that this biodiversity-rich mangrove forest provides. Through a convenient sampling technique, 421 tourist respondents were interviewed to assess their willingness to pay for the tourism services of the Sundarban, using the Zonal Travel Cost Method (ZTCM). The estimated annual economic contribution of tourism in the Sundarban mangroves to the Bangladesh economy is USD 53 million. The findings of this study showed that facilities for watching wildlife and walking inside the forest can increase the number of tourists in the SRF. The findings also show that the availability of information like forest maps, wildlife precautionary signs, and danger zones would increase the number of tourists as well. Thus, the government of Bangladesh should consider increasing visitor entry fees to fund improvements and to enhance the ecotourism potential of the Sundarban mangroves.

## INTRODUCTION

Ecotourism is a large and growing component of international tourism, which provides a means of earning foreign exchange and offers a less destructive use of resources (Honey, 2008). Additionally, ecotourism encourages the preservation of traditional customs, handicrafts and festivals that might otherwise be allowed to wane. In this regard, ecotourism creates civic pride (Buckley, 2009). Generally, protected areas are very important components of the ecotourism industry, as they occupy some of the most interesting landscapes. Ecotourism is increasingly becoming an important part of sustainable development because of the potential for contributing to local and national economic development, while providing incentives for nature and biodiversity

*Corresponding author. E-mail: dr.raihan.sarker@cu.ac.bd. Tel: +8801609479381.

conservation (Zacarias and Loyola, 2017).

The Sundarban Reserve Forest (SRF) of Bangladesh is a dynamic ecosystem that provides a variety of services, such as (1) provisioning (e.g., typical forest products, fisheries, etc.), (2) cultural (e.g., tourism, worship, educational research, etc.), (3) regulating (e.g., protection from cyclones, storm surges, floods, climate regulation, pollination, etc.), and (4) supporting (e.g., nursery and breeding ground of fish, nutrient cycling, habitat for biodiversity, etc.) services (Barbier, 2007; Giri et al., 2007; Kathiresan and Rajendran, 2005; Walters et al., 2008). Among the non-extractive services that mangrove forests offer are tourism services, which have great potential to benefit everyone in Bangladesh, its 163 million citizens and foreigners alike (BBS, 2019). The landscape of the SRF and its position as a tiger habitat make it a singular tourist attraction (Guha and Ghosh, 2011). The number of visitors to the Sundarban has been increasing day-by-day, and the region offers scenic beauty as well as wildlife-habitat, and traditional cultural and religious events, such as the Rash Mela and Ban Bibir Mela (that is, traditional worship of the local Hindu communities) (BFD, 2012; Hai and Chik, 2011).

There is a widespread notion that mangrove ecosystem valuation exercises might help decision makers which appreciate the value of the ecosystem services to the society and the anticipated cost of their imminent loss (Laurans and Mermet, 2014). Economic valuation in particular is often expected to be a useful tool to support conservation policy decisions and governance (Bateman et al., 2011). In this context, the USAID's CREL project with Winrock's JDR 3$^{rd}$ Scholars Program supported this research study to estimate the values (in monetary terms) of the tourism services of the SRF. A number of previous studies have been conducted on the economic valuation of the SRF, and these studies did not use primary data based on tourist surveys (Haque and Aich, 2014; Uddin et al., 2013). Therefore, the goal of this research was to provide an estimate of the economic valuation associated with the tourism services of the SRF of Bangladesh by using primary data collected through field surveys. Additionally, the specific objectives of the study were to understand the following: (i) the status and recreational behaviour of visitors, and (ii) the perceptions of visitors towards (a) the recreational activities and physical tourism services available in the SRF, and (b) the improvement of existing tourism services in the SRF.

## MATERIALS AND METHODS

### Information about the study site

The SRF is situated along the coastline of the Bay of Bengal in the south-western region (that is, Khulna division) of Bangladesh. It was declared a World Heritage Site by UNESCO in 1997. The total area of the forest is 603,000 ha and consists of three wildlife sanctuaries: Sundarban West (119,718.88 ha), Sundarban East (122,920.90 ha) and Sundarban South (75,310.30 ha) (BFD, 2021).

The present study was conducted in all existing forest ranges of the SRF, as defined by the Bangladesh Forest Department (BFD), such as (i) the Burigoalini Forest Range, (ii) the Khulna Forest Range, (iii) the Chandpai Forest Range, and (iv) the Sharankhola Forest Range (Figure 1).

### Questionnaire development, data collection, and analyses

Primary data was collected from the field directly through face-to-face interviews. Secondary data was collected from published books and journals and from unpublished sources like the reports of the BFD and information from different tour operators regarding the number of visitors and revenue earned from tourism. Data was collected from all tourist spots in the Sundarban Mangrove Forest from November, 2018 to April, 2019. A semi-structured questionnaire was used to collect information about the visitors' biographic information, visit preferences, reasons for preferring the Sundarban, selection of tour packages, opinions on existing tourism facilities and services, suggestions for improvement, and willingness to pay higher prices for improved quality of services provided by various tour operators.

During the study, a convenient sampling method was used in collecting primary data. Reaching the tourists on site at the tourist spots was convenient for the researchers. Usually, tourists visited in the Sundarban by ships provided by tour operators. In connection with this, research assistants for this study accompanied the tourists on these ships. The tourists were interviewed while returning from several Sundarban tourism spots. A total of 421 visitors (both local and foreign) were interviewed face-to-face. Day-visitors were also included, most of whom start from Karamjal. Thus, they were interviewed at that particular place.

Various Sundarban tour packages are offered by the tour operators. This study identified the packages through a Focus Group Discussion (FGD) with the established tour operators. All of the collected data (that is, primary data, secondary data and FGD information) was sorted for the purpose of quantitative and qualitative analyses. People who visit the Sundarban mangroves go for either tourism reasons or spiritual reasons. In this study, the responses of tourists only were considered.

The responses of spiritual pilgrims were excluded from the analysis. After processing and generating the data, visitors were categorized into seven major zones, based on their origin. Since there are seven divisions in Bangladesh, the tourists were categorized into one of these seven zones. The number of tourists in each zone was then divided by the total number of tourists to calculate the percentage share of each zone. The potential number of the visitors was measured based on the threshold level of income (minimum BDT 9158 or USD 117.42 to BDT 14092 or USD 180.69), which is the minimum income of the sampled visitors. The population of a zone having income above threshold level of income were considered as the potential number of tourists for that zone. Analyses were performed using SPSS version 20.0 (SPSS, Chicago, USA) and STATA 11. The study evaluated the differences in socio-economic condition, recreational behaviour, and tourists' attitude using one-way ANOVA and Chi-square ($\chi^2$) tests. The level of significance was set at p ≤ 0.05 and p ≤ 0.10.

### Calculation of the valuation of tourism service

The travel cost method (TCM) was applied for the valuation of tourism services in the Sundarban mangroves of Bangladesh. Gradually, TCM has become a widely used valuation technique for protected areas, national parks, sanctuaries, and forests. This method of valuing tourist spots, based on a tourist's willingness to pay, was developed by Harold Hotelling in 1949 when he used it to value U.S. national parks (Hotelling, 1949). There are two types of TCM, namely, the Individual Travel Cost Method (ITCM) and the Zonal Travel Cost Method (ZTCM). The ITCM is used when data is collected from individual tourist levels, and the ZTCM is used when data is collected from secondary sources. Defining the travel zones, the ZTCM is used for an identical recreational site. If secondary data sources such as recreational permits or fees are limited, the ZTCM is very useful (Loomis et al., 2009). In this study, the proportion of primary data was used for each zone to convert the secondary data by those proportions in order to use the ZTCM in estimating the value of tourism services of the Sundarban mangroves. The ZTCM is considered as one of the simplest and least-expensive methods for valuation of tourism service (Saraj et al., 2009). The main benefit of this method is the inclusion of the cost of time and travel that people incur for a visit, which exhibits as the price of travel to a site. Thus, ZTCM helps to estimate the willingness of tourists to pay based on their number of visits as a demand schedule with different prices. It also assumes that the cost for one individual to visit a recreation location from a specific zone is the same for all other individuals from that same zone (Emiriya, 2013), although there is heterogeneity among the population and variation in travel cost.

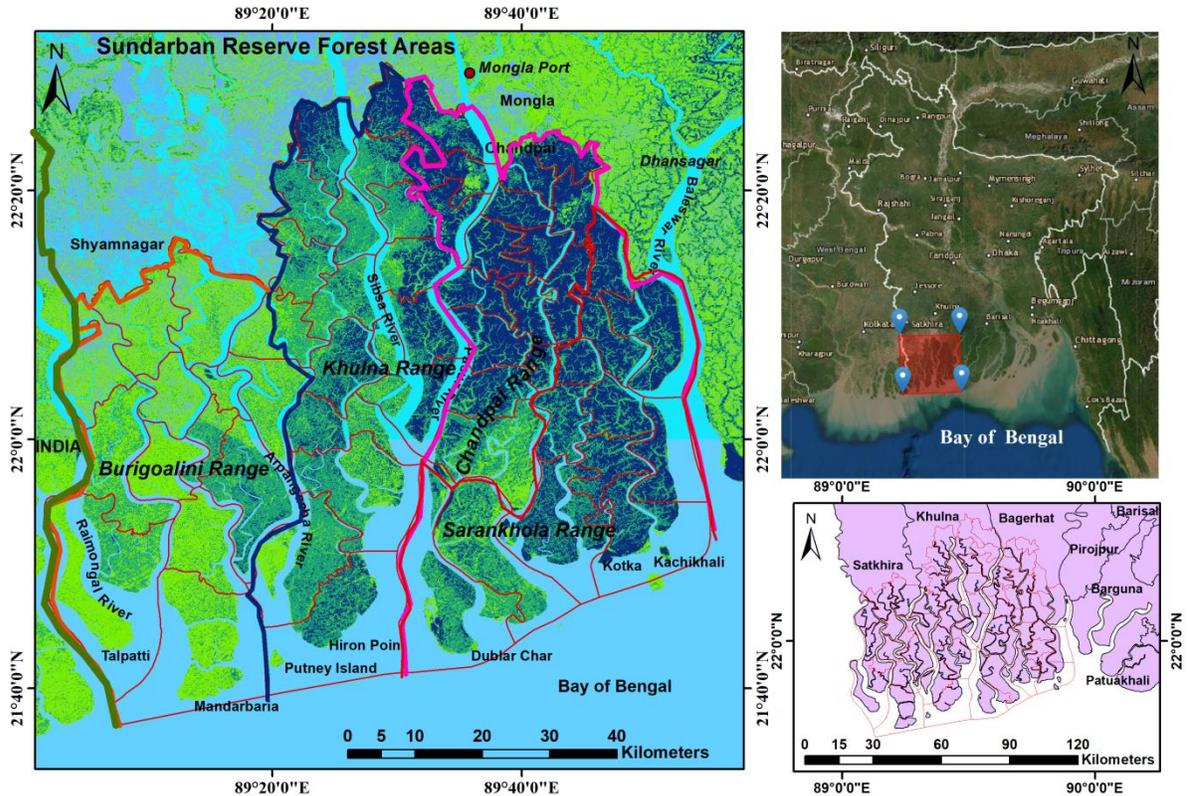

**Figure 1.** Location of the study area. Study sites have four forest range: Burigoalini Forest Range, Khulna Forest Range, Chandpai Forest Range and Sarankhola Forest Range shown in left panel. In this panel, base image used NASA SRTM DEM v3 acquisition from USGS Earth Explorer website: Image resolution: 1-Arc seconds: 30 meters; In Right Panel: Top image: Google Earth Pro view highlight SRF areas with transparent red box; and Bottom right: SRF location shown over District Map of Bangladesh as zoom view. The Sundarban Forest Compartment boundary and district boundary shape files are taken from Bangladesh Forest Department for JRD 3rd Scholar Project and DIVA-GIS web platform (https://www.diva-gis.org/gdata), respectively.

To get the actual number of visitors of different zones, the sample proportion (primary data collected during 2018-2019) was used to divide the actual number of visitors for 2018 into seven divisions. Then, these estimated numbers of visitors for 2018, categorized into seven zones, were used to estimate the zone-wise travel cost. The total number of visitors in 2018 for each zone was divided by the potential number of visitors of that zone to calculate the zone's visitation rate ($V_i$). The potential number of visitors refers the population with enough income to visit a place. The income distribution of the respondents was used to identify the potential number of visitors. From the income distribution of the respondents, which is found in the descriptive statistics, minimum income level of the respondents of each zone was identified. The threshold income level of a particular zone indicates people in that zone with the minimum income level or above, as these are the people who can afford to visit the Sundarban mangrove forest. For instance, the income of the top 5% of the population in Dhaka Division remains above that division's threshold level of income, and thus those people are considered as potential visitors to the Sundarban. The potential population of tourists was identified from the *'Household Income Expenditure Survey 2011'* (BBS, 2011) and categorized by zones following the administrative boundaries of the seven divisions in Bangladesh.

The number of visitors to a recreational area and the distance visitors travelled is used in the ZTCM to calculate the *'price'* that the visitors pay to visit the site (King and Mazzoatta, 2000). The ZTCM counts how many people travel from different distances or zones from the park and estimates the cost of the various distances. By adding the number of people and their associated travel costs, the total price that tourists pay to visit the site can be calculated. Thus, it is important to know the visitors' zone of origin to determine the visitation rate. This visitation rate considers cost of travel from the origin to the recreation site, and the income or demographic characteristics of the respondents (Das, 2013).

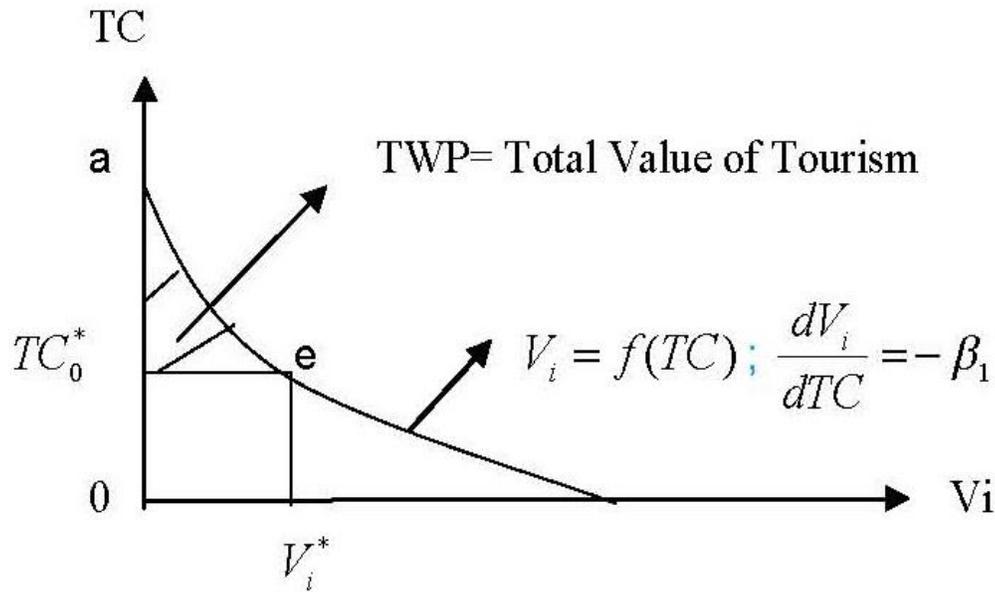

**Figure 2.** Visit-generating function ($V_i$).

The visitation rate ($V_i$) was estimated as the visit-generating function (*VGF*) on costs of travel and others demographic variables (Nobi and Nahar, 2013). The visit-generating function was considered as the demand function for the visit, where *VGF* is the dependent variable and the travel cost (that is, price of visit) is the core explanatory variable along with other explanatory variables, such as whether the tourist is alone or in a group, whether the tourists uses at least one way by air or not, whether the tourist is from Khulna or some other division of Bangladesh, and whether the tourist is part of a package tour offered by tour operators. After the estimation, the price elasticity was calculated from the demand function for visitation to calculate the tourists' willingness to pay (WTP) for different zones. The total willingness to pay was then calculated by adding the WTP for different zones, which resembles the total value of tourism service of the SRF.

The visitation rate ($V_i$) of the visit-generating function is inversely related to the travel costs (Figure 2). This means that as the travel costs decrease the demand for tourism increases. The area under the *VGF* (that is, demand function for visit) shows the total economic value of tourism service. Subtracting the total costs of travel (that is, total costs = average travel costs × number of visitors) from the total value of tourism provides the net value of tourism, which is the consumer surplus (*CS*) and the contribution of tourism service to the national economy. According to Figure 2, area $aeTCo$ is the net value for each zone in terms of willingness to pay. Finally, the total value of the tourism service of the Sundarban mangroves was calculated by adding the consumer surpluses of all zones.

**The econometric model of the zonal travel cost method (ZTCM)**

Adopted from Clawson (1959) and with a modification according to Torres-Ortega et al. (2018), the econometric model of the ZTCM is as follows:

$$V_i = f(TCost_i, Alone, Air, Khln, Package) \quad (1)$$

$$V_i = \alpha + \beta_1 TCost_i + \beta_2 Alone + \beta_3 Air + \beta_4 Khln + \beta_5 Package + \varepsilon) \quad (2)$$

$$V_i = \frac{Total\ number\ of\ visit}{Total\ population}$$

where $V_i$ stands for visitation rate, which is the dependent variable that is calculated by dividing the zonal number of visitors by their total population. The explanatory variables are *TCost*, the travel cost of zone *i*; *Alone*, a dummy variable with a value of either 1 (the specific visitor is alone) or 0 (the visitor is in a group); *Air*, a mode of travel used by visitors from any destination to Jessore (the nearest airport to the SRF); *Khln*, a dummy variable representing zone, where 1 indicates the visitor is from Khulna Division and 0 indicates all other divisions; and *Package*, a dummy variable where the value 1 indicates the visitor is in a package tour and 0 indicates no package tour. $\beta_0$ is a constant variable and $\varepsilon$ is the residual. It is notable that $V_i$ is a dependent variable, while $TCOST_i$, *Alone*, *Air*, *Khln* and *Package* are independent variables in the model estimation. Holding other things constant, *VGF* ($V_i$) is inversely related with travel costs of the zones. Thus, the slope coefficient of travel cost ($\beta_1$) of this *VGF* is used to calculate the value of tourism service; the *CS* of tourism in the Sundarban mangroves.

**Theoretical framework for calculating the areas of tourism locations**

After calculating the tourism value, it was divided by the total number of visitors to get consumer surplus per visit. It is also divided by the total area of different tourism locations that the tourists visit in order to determine the tourism value per hectare of the SRF. In every tourist spot, visitors usually go to one side of the river in the Sundarban. Table 1 depicts the distance in km that the visitors walk, which has been calculated by using GIS (that is,

**Table 1.** Tourist coverage at some selected tourist spots in the SRF of Bangladesh.

| Tourist spots | Compartment (Tourist spot) no. | Distance (km) | Area in half of circle (km$^2$) | Area (ha) |
|---|---|---|---|---|
| Katka (including Jamtala sea beach) | 7 | 6 | 56.6 | 5654.9 |
| Kachikhali | 6 | 2 | 6.3 | 628.3 |
| Harbaria | 26 | 4 | 25.1 | 2513.3 |
| Kalagasia | 46 | 3.5 | 19.2 | 1924.2 |
| Karamjol | 31 | 5 | 39.3 | 3926.9 |
| Nilkamal | 44 | 2 | 6.3 | 628.3 |
| Dubla, Meherali, Alorkol, and Majherchar | 45 | 15 | 353.4 | 35342.9 |
| Total | - | 37.5 | 506.2 | 50618.9 |

$\Pi = 3.14159$ and 1 km$^2$ = 100 ha.

*Geographic Information System*) technology where the visitors walk around. While measuring the favourite area for each tourist spot, one side of the river where visitors usually walk around in each tourist spot was considered. The estimated area for each of the seven tourism locations is shown in Table 1. Finally, to estimate the value per hectare of tourism service in the Sundarban mangroves, the total estimated value of tourism (in terms of WTP) was divided by the total area of Sundarban mangroves visited by tourists.

## RESULTS

### Socio-economic and demographic profiles of visitors/tourists

Among the visitors (n = 421), the majority were (n = 369, 87.6%) local, while the remaining (n = 52, 12.4%) were foreigners, who came from the outside of the country, especially from USA (4%), the UK (2.1%), and Germany (1.9%). During the study, it was observed that a considerable number of foreign visitors also came from a variety of other countries, that is, Australia, Brazil, Canada, Croatia, France, India, Nepal, Netherland, Singapore, Sri Lanka, and Italy. Among local visitors (n = 369), the highest numbers were estimated from the Dhaka division (41.2%) followed by Khulna division (39.8%), with the remaining from the Rajshahi (6.8%), Chittagong (6.2%), Rangpur (3.0%), Barisal (2.2%) and Sylhet (0.8%) divisions. The proportion of males was higher than females, and visitors mostly included young, middle aged, and elderly (Table 2).

Table 2 shows the socio-economic and demographic profiles of visitors. More than two-thirds of the respondents were highly educated (having graduate degree or above), and the number of these kinds of visitors was greater among foreigners than among local visitors. This study also identified the marital status of the tourists. The majority of local visitors were married, with the remaining being unmarried, divorced or widowed. The profession of the respondents was also varied. The majority of tourists were students. The average household size of the visitors was 4.5 persons (± 2.49), and the household size significantly varied between the types of visitors (*that is,* local: 4.83 ± 2.45 persons per household, foreigner: 2.21 ± 1.40 persons per household, F = 51.44, df = 1, and p = 0.0001). The average income of local visitors was approximately BDT 66,000/month or USD 846.26.

### Recreational behaviour of visitors

More than four-fifths of the visitors (n = 421, 81.2%) said that they visited the Sundarban mangrove forest for the first time, while 18.8% were the repeat visitors, and their responses varied significantly between the types of respondents ($\chi^2$ = 6.57, df = 1, p = 0.01). Among the repeat visitors, 20.6% of locals and 5.8% of foreigners were visiting the area for the second time. About 90.5% of the visitors reported that they came under a tour package, and their proportion varied significantly between the types of visitors ($\chi^2$ = 3.96, df = 1, p = 0.047). More than two-thirds of the visitors (67.9%) said that they learned about the SRF from their family, friends, and relatives, while the remaining visitors mentioned media (9.7%), organizational websites (7.6%), social networks (7.6%), advertisements of tour operators (5.7%), government websites (1.0%), and newspapers, books, etc. (0.5%). Their responses varied significantly between the types of visitors ($\chi^2$ = 41.68, df = 6, p = 0.0001). This study also revealed that a significant number of visitors (87.4%) travelled in groups, while 12.6% were visiting alone. Their responses varied significantly between the types of visitors ($\chi^2$ = 30.92, df = 1, p = 0.0001).

More than half of the visitors (59.1%) mentioned that the Sundarban mangrove forest areas were the only destination within the region they visited on this trip. A majority of the visitors (84.1%) reported that they visited the Sundarban for recreation, while 12.4% visited for spiritual purposes, 3.5% came for study purpose, and 1% for business purposes. The purpose of visit varied significantly between the types of visitors ($\chi^2$ = 15.14, df =

Table 2. Socio-economic data, obtained from interviews of visitors in Sundarban included in the study and $\chi^2$ tests of independence between types of visitors.

| Socio-economic variables | | Types of visitor (%) | | Total (n = 421; %) | Statistics | | |
|---|---|---|---|---|---|---|---|
| | | Foreign tourist (n = 52) | Local tourist (n = 369) | | $\chi^2$ | df | p |
| Sex | Female | 30.80 | 10.00 | 12.60 | 17.82 | 1 | 0.0001 |
| | Male | 69.20 | 90.00 | 87.40 | | | |
| Age | Youth (18-30 years) | 28.0 | 37.8 | 36.6 | 16.26 | 2 | 0.0001 |
| | Middle age (31 to 50 years) | 38.0 | 49.9 | 48.4 | | | |
| | Old (above 50 years) | 34.0 | 12.3 | 14.9 | | | |
| Education | Below primary level/No. schooling | 0.00 | 2.20 | 1.90 | 9.67 | 3 | 0.022 |
| | Up to primary | 1.90 | 0.50 | 0.70 | | | |
| | Up to secondary | 11.50 | 29.00 | 26.80 | | | |
| | Graduate or above | 86.50 | 68.30 | 70.50 | | | |
| Marital status | Unmarried | 40.40 | 31.20 | 32.30 | 31.54 | 3 | 0.0001 |
| | Married | 51.90 | 68.80 | 66.70 | | | |
| | Divorced | 5.80 | 0.00 | 0.70 | | | |
| | Widow | 1.90 | 0.00 | 0.20 | | | |
| Occupation | Agriculture | 0.00 | 1.90 | 1.70 | 24.91 | 8 | 0.002 |
| | Business | 19.20 | 24.70 | 24.00 | | | |
| | Housewife | 0.00 | 2.20 | 1.90 | | | |
| | Journalist | 1.90 | 0.50 | 0.70 | | | |
| | Researcher | 3.80 | 0.30 | 0.70 | | | |
| | Student | 25.00 | 44.20 | 41.80 | | | |
| | Teacher | 7.70 | 4.30 | 4.80 | | | |
| | Technical profession | 21.20 | 10.60 | 11.90 | | | |
| | Others (GOs & NGOs) | 21.20 | 11.40 | 12.60 | | | |

3, p = 0.002). The visitors used buses (62.0%), boats (26.4%), and a combination of air, bus, and boat (11.6%) to travel to the Sundarban, and their responses varied significantly between the types of visitors ($\chi^2$ = 393.61, df = 2, p = 0.0001). More than 85% of visitors travelled to the Sundarban for a short duration only (3 to 7 days), and 63.7% of visitors covered tour expenses with their own income; 27.9% of visitors reported sponsorship, and 6.8% reported support received from family and friends.

Moreover, more than one-fifths of the respondents (21.4%) expressed dissatisfaction towards the overall recreational arrangements in the Sundarban areas, and their perceptions varied significantly between the types of visitors ($\chi^2$ = 8.11, df = 1, p = 0.0001). When asked 'would you be willing to pay to conserve the environment and endangered species in the SRF?' more than half of the visitors (51.3%) responded "yes." Among them, the proportion was higher for local (53.1%) than foreign (38.5%) visitors, and their responses varied significantly between the types of respondents ($\chi^2$ = 3.92, df = 1, p = 0.05). More than half of the visitors (51.3%) wanted to see improvements in facilities required for watching wildlife (specifically birds, deer, and tigers), while 96.4% wanted improvements for walking facilities for going deeper into the forest, 45.6 and 49.4% requested improvement and facilities should increase to watch deer and bird, respectively, and 7.4% requested improved facilities for religious functions. Approximately 47% of visitors wanted to see improvement in information services (e.g., forest maps, wildlife precautionary signs, danger zones, etc.). Among respondents who requested improvements in the recreation and information facilities, the proportion was significantly higher for local (76.2%) than foreign (44.2%, $\chi^2$ = 23.14, df = 1, p = 0.0001) visitors. However, again when we asked, 'what is your opinion about forest entry fee?' in response to this, more than half (52.7%) of the respondents said "reasonable"; and among them, the proportion was considerably higher for local (55.0%) than foreign (36.5%) visitors. Among them, the remaining respondents, 21.1% of the

**Table 3.** Estimated result of the visit-generating function.

| Variable | Coefficients | Robust SE | T | P > t | [95% Conf. Interval] |
|---|---|---|---|---|---|
| $TCOST_i$ | -0.00016* | 6.5 | -25.50 | 0.03 | - 0.0003 - 1.8 |
| Alone | 487.94** | 328.2 | 1.49 | 0.16 | - 234.36 + 1210.24 |
| Air | -23.15** | 16.2 | -1.43 | 0.18 | - 58.86 + 12.56 |
| Khln | 58.31** | 40.4 | 1.44 | 0.18 | - 30.66 + 147.29 |
| Package | 34.99*** | 26.5 | 1.32 | 0.21 | - 23.44 + 93.43 |
| Constant | 1.15*** | 20.6 | 0.06 | 0.96 | - 44.14 + 46.45 |

Linear regression: Number of observation= 17, F (5, 11) = 3.70, Prob> F = 0.0328, R-squared = 0.6030, Root MSE = 45.969 *Indicates the level of significance below 5% level, ** indicates the level of significance up to 10%.
Source: Estimated from the sample data.

respondents mentioned "less", 11.2% mentioned very less, 3.8% very high, and 11.2% high. Approximately 70% of visitors stated that the government should allocate more money to provide improved services, but also agreed to pay higher entry fees if government resources were not available for improvement of the facilities. Among the respondents, the proportion was significantly higher for foreigner (84.6%) than local (67.5%, $\chi^2$ = 6.32, df = 1, p = 0.01) visitors.

**Estimation of the visit generating function**

The estimated results are shown in Table 3. The visitation rate (the demand to visit) is negatively influenced by the travel cost from different zones, which is statistically significant (Table 3). However, the coefficients of the other variables of the visit generating function are insignificant. The result shows that the targeted variable travel cost ($TCost_i$) of different zones is inversely related with the visitation rate ($V_i$), which meets the properties of a travel demand function. All independent variables explain 60.3% of the variation in visitation rate with 17 observations. Though, the actual number of the sample observations is 421 but converting to different zones the observation turned to 17. Thus the number of observations here represents the zones which have been restructured by the mode of transports of the visitors of the seven divisions. The F-value is significant below to 5% level of significance refers that the model is a good fit.

**Value of tourism services of the Sundarban mangroves**

The slope coefficient of travel cost from visits generating a function was used to calculate the value of tourism service in the Sundarban mangroves of Bangladesh. The value of the tourism service in the SRF for different zones and their aggregation is summarized in Table 4. The analysis revealed that tourism as a whole in the SRF contributes about USD 53.14 million/year (Table 4) to the economy of Bangladesh.The total value of tourism was divided by the total visits (in the year 2018) to estimate the value of tourism per visit, and it was calculated as USD 577. The tourism service/hectare was also estimated by dividing the total value of tourism services by the total areas of seven tourism locations (that is, 50,619 ha), which provides USD 1,050/ha/year. After calculating the value per hectare of the tourism service in the SRF, the value of various tourism locations was calculated in order to determine the importance of individual tourism locations. It was revealed that a few tourist spots (that is, Dubla, Meherali, Alorkol, and Majherchar) constituted the highest value of tourism, compared to the other tourist spots (that is, Kotka, Karamjol, Harbaria, Kalagasia, Kachikhali and Nilkamal). The estimated value of each of the tourist spots is shown in Table 5.

**DISCUSSION**

The SRF contributes about USD 53.14 million/year to the national economy of Bangladesh, and it offers tremendous opportunities for developing ecotourism in terms of tourist attractions, economic benefits, employment, and ecosystem conservation (Alam et al., 2010; Hasan, 2012; Islam, 2008; Siddiqi, 2001). Haque and Aich (2014) studied the economic valuation of the ecosystem services of the SRF in Bangladesh. Applying the Delphi approach, the researchers considered nine support services, seven regulating services, five provisioning services, and three cultural services. Their estimate showed that the total value per hectare of the land of the Sundarban varied from USD105 to USD840 per year. Uddin et al. (2013) estimated the value of cultural services of the Sundarban based on the revenue collected by the Bangladesh Forest Department (BFD) only. According to their study, cultural services (that is

**Table 4.** Calculation of the tourism values of SRF for different zones.

| Name of zones | Zonal travel cost (BDT) | Actual visits in 2018/10000 | Potential visitors in 2018/10,000 | Choke price* (BDT) | TWTP[1] (million BDT) |
|---|---|---|---|---|---|
| Barisal | 13,203.4 | 19.3 | 42.4 | 133,787.4 | 60.3 |
| Chittagong | 26,517.8 | 16.1 | 135.8 | 127,014.4 | 167.9 |
| Dhaka | 14,018.4 | 34.9 | 227.8 | 232,247.8 | 980.8 |
| Khulna | 11,699.2 | 102.0 | 76.9 | 648,468.8 | 2590.5 |
| Rajshahi | 3,772.4 | 27.7 | 90.1 | 176,831.7 | 225.7 |
| Rangpur | 8,188.6 | 16.7 | 77.2 | 112,412.9 | 77.7 |
| Sylhet | 162,727.3 | 5.1 | 45.9 | 194,581.6 | 41.9 |
| Total value | BDT 4,144.7 million (or USD 53.14 million) | | | | |
| Mean CS | BDT 45000 (or USD 577) | | | | |
| Value/ha | BDT 81880.3 or USD 1,049.75 | | | | |

The calculation is based on estimated coefficient of VGF. *Choke price is the travel cost at which demand for visit from each of the zone comes to zero. It is calculated by dividing the intercept coefficient of the visit generating function by its slope coefficients. [1]Total Willingness to pay.
Source: Estimated from the sample data.

**Table 5.** Estimated value of tourism service of the different tourist spots of the SRF.

| Name of the tourist spot (compartment no.) | Area (ha) | Value per year (million USD) |
|---|---|---|
| Katka included Jamtala sea beach (7) | 5654.9 | 5.9 |
| Kachikhali (6) | 628.3 | 0.7 |
| Harbaria (26) | 2513.3 | 2.6 |
| Kalagasia (46) | 1924.2 | 2.0 |
| Karamjol (31) | 3926.9 | 4.1 |
| Nilkamal (44) | 628.3 | 0.7 |
| Dubla, Meherali, Alorkol, and Majherchar (45) | 35342.9 | 37.1 |
| Total | 50618.9 | 53.1 |

tourism) contributed on average USD42,000 per year from Fiscal Year 2001-2002 to 2009-2010. However, the Government of Bangladesh has already developed a travel guideline for the Sundarban mangroves; there is an urgent need to develop policy guidelines for the sustainable development of eco-tourism in the Sundarban. In this study, it is revealed that the higher proportion of visitors are literate and earn more than the average income level in the country based on 2010 Household Survey data (HIES, 2010). Since tourism expenditures have been found to be directly linked with family income (Abbruzzo et al., 2014), household income has a positive impact on recreational behavior (Landry et al., 2012). It has been shown that visitors can afford to and are willing to pay higher rates to enjoy the amenity of the Sundarban. Visitors with higher incomes may be willing to pay more if the recreational quality of a park improves (Sarker et al., 2017). However, the added revenue generated from the forest service can be used to address some of the degradation challenges that the forest is facing (Brander et al., 2012, Brougham and Butler, 1981; Cavus and Tanrisevdi, 2002).

In this study, it is revealed that among the repeat visitors (we asked whether it is their first/second visit), the proportion of both locals and foreigners was very low, and almost half of them wanted to see the improvement (based on a question on what improvement they want to see) in information services such as forest maps, wildlife precautionary signs, danger zones, etc. Unfortunately, the government and private sector organizations have not yet developed well-organized, informative, and educational guides, signs, materials, or websites on the Sundarban tourism for visitors (BIDS, 2010). This is one of the major drawbacks for the growth of tourism in the Sundarban mangroves. Information regarding the natural and cultural significance of the Sundarban mangroves could motivate tourists to become aware of their ecological footprints, garner greater appreciation for the unique ecosystem, and enhance their experience (Budeanu, 2007). Trained tour guides could organize and lead tourists in more efficient ways to enhance their interest for further travel to the Sundarban. In this regard, the government should establish an "Ecotourism Training Centre" at Khulna and the surrounding areas of the Sundarban to ensure better customer service by producing skilled eco-tourism guides and creating employment opportunities in eco-tourism enterprises. The training centre should also provide orientation training to visitors before they enter the forest, regarding laws and regulations with respect to nature, environmental philosophies, biodiversity, conservation issues, etc., of the SRF.

However, this study shows that more than 20% of visitors expressed dissatisfaction towards the overall recreational arrangements in the Sundarban mangroves of Bangladesh. Visitors would like to see improvements in the facilities for watching tigers, deer, and birds, as well as improved walking facilities for going deeper into the forest. In this regard, more than two-thirds of the visitors also agreed to pay higher entry fees if government resources were not available for improving the facilities. Thus, the government of Bangladesh should consider increasing visitor entry fees to improve and enhance the eco-tourism potential of the Sundarban mangroves.

Finally, the BFD should facilitate and engage with landscape communities and stakeholders to develop a participatory community-based ecotourism strategy to ensure long-term local community benefit-sharing and promotion of activities run by local communities. A system should be developed in which increased ecotourism revenues collected by the BFD would be used for the protection and conservation of the Sundarban's biodiversity, local livelihoods, and overall development. Further, the co-management committees can also be engaged to support the Sundarban's ecotourism activities, which will ultimately help to achieve the specific UN sustainable development goals (SDGs) 2030.

## Conclusion

The unique landscape of the SRF and its position as a tiger habitat make it a singular attraction for tourists. The direct economic contribution of tourism in the Sundarban is the entry fee of visitors charged by the BFD, along with other expenses related to travel, food, and accommodation charged by different service providers. Therefore, cultural services such as ecotourism contribute a significant amount of economic benefit to the national economy of Bangladesh. Moreover, the government must develop a system by which visitor entry fees, collected by the BFD, will be used for the protection and conservation of the SRF and local livelihoods, and for overall aesthetic development. Finally, the BFD should facilitate and engage with landscape communities and stakeholders to develop a participatory community-based ecotourism strategy to ensure long-term local community benefit-sharing and promotion of activities run by local communities. Though this research provides an in-depth visualization of ecotourism potential of the SRF in Bangladesh, which expected further research needs to be carried out in the coming days, especially when its importance will arise as demanded by the forest department.

## CONFLICT OF INTERESTS

The authors have not declared any conflict of interests.

## ACKNOWLEDGEMENT

This study was carried out under the financial assistance provided by the JDR-Winrock-USAID.